\documentclass[aps,prb,amsmath,amssymb,reprint]{revtex4-1}
\pdfoutput=1
\usepackage{graphicx}
\usepackage[utf8]{inputenc}
\usepackage{hyperref}

\begin{document}

\title{Static magnetic proximity effect in Pt/Ni$_{1-x}$Fe$_x$ bilayers investigated by x-ray resonant magnetic reflectivity}

\author{C.~Klewe$^1$, T.~Kuschel$^1$, J.-M.~Schmalhorst$^1$, F.~Bertram$^2$, O. Kuschel$^3$, J.~Wollschl\"ager$^3$, J.~Strempfer$^2$, M.~Meinert$^1$, G.~Reiss$^1$\email{Electronic mail: cklewe@physik.uni-bielefeld.de}}
\affiliation{$^1$Center for Spinelectronic Materials and Devices, Department of Physics, Bielefeld University, Universit\"atsstra\ss e 25, 33615 Bielefeld, Germany\\
$^2$ Deutsches Elektronen-Synchrotron DESY, Notkestra\ss e 85, 22607 Hamburg, Germany\\
$^3$ Fachbereich Physik, Universit\"at Osnabr\"uck, Barbarastra\ss e 7, 49069 Osnabr\"uck, Germany}

\date{\today}

\begin{abstract}
 
We present x-ray resonant magnetic reflectivity (XRMR) as a very sensitive tool to detect proximity induced interface spin polarization in Pt/Fe, Pt/Ni$_{33}$Fe$_{67}$, Pt/Ni$_{81}$Fe$_{19}$ (permalloy), and Pt/Ni bilayers. We demonstrate that a detailed analysis of the reflected x-ray intensity gives insight in the spatial distribution of the spin polarization of a non-magnetic metal across the interface to a ferromagnetic layer. The evaluation of the experimental results with simulations based on optical data from {\it ab initio} calculations provides the induced magnetic moment per Pt atom in the spin polarized volume adjacent to the ferromagnet. We find the largest spin polarization in Pt/Fe and a much smaller magnetic proximity effect in Pt/Ni. Additional XRMR experiments with varying photon energy are in good agreement with the theoretical predictions for the energy dependence of the magnetooptic parameters and allow identifying the optical dispersion $\delta$ and absorption $\beta$ across the Pt L$_3$-absorption edge.

\end{abstract}

\maketitle

\section{Introduction}
The generation and detection of pure spin currents play an essential role in spintronics\cite{Wolf2001} and spincaloritronics\cite{Bauer2012}.
A variety of spintronic effects, like spin pumping\cite{Tserkovnyak2002, Mosendz2010, Czeschka11}, the spin Hall effect\cite{Hirsch99, Hoffmann2013}, and the recently reported spin Hall magnetoresistance\cite{Nakayama2013, Althammer2013, Chen2013} are closely related with such spin currents.
In spincaloric transport, the longitudinal spin Seebeck effect (LSSE)\cite{Uchida2010}, where a spin current is generated in a ferromagnetic material parallel to an out-of-plane temperature gradient, is the most popular example associated with pure spin currents. 

However, the detection of pure spin currents still remains challenging.
The most common approach is to use a non-magnetic metal (NM) with a large spin orbit coupling and convert the spin current into a voltage via the inverse spin Hall effect (ISHE)\cite{Saitoh06}. 
In particular, noble metals like Pd\cite{Tang13}, Au\cite{Seki08}, and Pt\cite{Liu2011} as well as the 5d transition metals W\cite{Pai2012} and Ta\cite{Liu2012} in their $\beta$-phase have proven to be suitable NM spin detectors, due to their large spin Hall angle. 

So far, in most publications Pt has been chosen as a spin detector material. 
Though, static magnetic proximity effects can occur at the interface of Pt to a ferro(i)magnetic layer (FM) due to its close vicinity to the Stoner criterion. 
Such a spin polarization in Pt can give rise to additional parasitic charge current effects hampering the evaluation of the ISHE voltage. 
Prominent examples for such parasitic effects that can deteriorate data interpreted as spin Seebeck effect\cite{Huang12, Meier13, Schmid13, Meier14} are the anomalous and the planar Nernst effect. 
Further, it has been shown that the spin Hall effect in a NM adjacent to a FM can be significantly reduced in the presence of magnetic proximity effects\cite{Zhang2015}.
Therefore, a detailed investigation of the magnetic properties of the NM/FM interface is essential for the distinction of ISHE voltages generated by pure spin- or parasitic charge-currents. 

\begin{figure}[t]
	\centering
		\includegraphics[width=8.5cm]{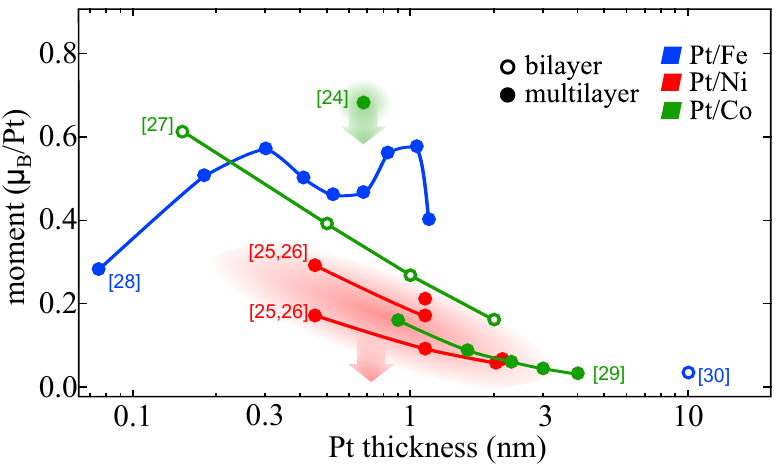}
	\caption{(Color online) Overview on publications investigating magnetic proximity effects in Pt/FM bi- and multilayers by XMCD. Series from the same publication or comparable sample systems are connected with solid lines. The references are included in the graph. Note that the moments from Refs.[\cite{Wilhelm03, Wilhelm2000, Poulopoulos01}] were measured at 10K, while the other values have been recorded at room temperature (RT). Therefore, these results show a larger XMCD response and should be lower at RT, as indicated by the shaded areas.}	
	\label{fig:Proximity_Pt}
\end{figure}

The most common technique to element selectively investigate the magnetic properties of thin films is x-ray magnetic circular dichroism (XMCD)\cite{Schuetz87}.
Here the absorption spectra across the L$_{2,3}$ absorption edge are recorded using circularly polarized x-rays, both for an applied magnetic field in parallel and antiparallel orientation to the in-plane projection of the incident beam. This allows to determine the absolute moment per atom of each element.
For more than two decades, there have been publications addressing the subject of spin polarizations in NMs induced by magnetic proximity using this technique.
Fig.\ref{fig:Proximity_Pt} gives an overview on publications discussing this topic for Pt in proximity to metallic FMs investigated with XMCD.
Unfortunately, for investigations of the interface spin polarization, as essential for evaluating magnetic proximity effects, the XMCD yield strongly depends on the film thickness, since the whole film volume contributes to the absorption\cite{Suzuki05, Tomaz97, Schwickert98}. Therefore, a small interface-to-volume ratio can hide out contributions from the interface. Thus, data on the interface spin polarization from XMCD can become difficult to analyze for films thicker than a few nm. 
Therefore, most of the displayed XMCD investigations on magnetic proximity in Pt were carried out on films with very small thicknesses, while larger thicknesses lead to a vanishing moment\cite{Rueegg91, Gepraegs12}. This is also emphasized in Fig.\ref{fig:Proximity_Pt}.
Additionally, Ederer et al. have stated, that the sum rule analysis carried out to extract the magnetic moments from XMCD spectra might result in large uncertainties and gives at best semiquantitative results for interface spin polarizations in Pt\cite{Ederer02}.

A presently less well established technique to detect the element resolved magnetic properties of layer systems with a focus on their interfaces is XRMR\cite{Geissler01, Macke14}. 
Contrary to XMCD in absorption, the magnetic circular dichroism in XRMR is observable in the interference of the light reflected from the interfaces. Therefore, this method is independent of the layer thickness as the main contributions to the signal originate from the surface and the interfaces. 
In a previous work we discussed XRMR experiments on Pt/Fe and Pt/NiFe$_2$O$_4$(NFO) bilayers\cite{Kuschel14} in order to evaluate parasitic contributions in LSSE studies on this system\cite{Meier13}.
Besides the absence of magnetic proximity in Pt/NFO we were able to show that the asymmetry ratio of the XRMR data is not depending on the Pt thickness in a range between 1.8\,nm and 20\,nm. This aspect represents a major advantage of XRMR over the strongly thickness dependent XMCD (see Fig.\ref{fig:Proximity_Pt}).
Additional to these benefits, XRMR also provides information on the spatial distribution of the magnetic moments across the interface.

A basic understanding of the physics behind XRMR can be obtained by considering the optical properties of the investigated material.
The complex refractive index of a material is given by $n = 1- \delta + i \beta$, where the real part $\delta$ is the dispersion coefficient and the imaginary part $\beta$ is the absorption coefficient, which are connected via the Kramers-Kronig relation. In magnetic materials $\delta$ and $\beta$ vary by a fraction $\pm \Delta \delta$ and  $\pm \Delta \beta$, respectively, depending on the orientation of the magnetization relative to the incident beam. These variations of $\delta$ and $\beta$ are most pronounced at photon energies close to the absorption edge of a material.
A detailed description of the theoretical background of XRMR is given in Ref.\cite{Macke14}.

In this work we report results for Pt/Fe, Pt/Ni$_{33}$Fe$_{67}$, Pt/Ni$_{81}$Fe$_{19}$, and Pt/Ni bilayers. 
We present three different approaches to simulate the magnetooptic profiles of $\Delta \delta$ and $\Delta \beta$ as a function of the coordinate perpendicular to the interface to optimize the fit routine. Then, we extract the spatial distribution of the spin polarization in Pt and evaluate the best fitting model.

In a second step we perform XRMR measurements for different photon energies on Pt/Fe in order to exclude any contributions from other absorption edges and to ascribe the observed asymmetry ratios to the dichroism unequivocally. 
A comparison of the acquired magnetooptic coefficients with theoretical predictions from {\it ab initio} calculations allows extracting quantitative values for the induced moments per Pt-atom in the effective spin polarized volume.

In the last part we present XRMR studies on Pt/FM bilayer systems with different FMs and discuss the induced magnetic moments in Pt regarding the Fe content in each FM.

\section{Experimental and theoretical details}

All bilayer systems were fabricated by dc magnetron sputter-deposition on (001) oriented MgAl$_2$O$_4$ (MAO) substrates at room temperature (RT). The Ar process pressure was $2\cdot 10^{-3}\,$mbar.

The XRMR measurements were carried out at the resonant scattering and diffraction beamline P09 of the third generation synchrotron PETRA III at DESY (Hamburg, Germany)\cite{Strempfer13}. 
For our studies a 6-circle diffractometer was used to perform XRR scans in a $\theta$-$2\theta$ scattering geometry. 

Except for the studies at varying photon energies, the XRMR data were collected at a fixed energy close to the peak of the Pt L$_{3}$ absorption edge. 
At this photon energy the x-ray reflectivity (XRR) scans were collected using circularly polarized x-rays, while a magnetic field was switched between parallel and antiparallel orientation to the in-plane projection of the incident beam at every incidence angle.
A four coils electromagnet was used to apply the external magnetic field.
The maximum applied field was $\pm 85\,$mT. The degree of circular polarization was $(99 \pm 1)\%$ for left and right circular polarization as determined from a polarization analysis with a Au(111) analyzer crystal. All measurements were carried out at RT with left circular polarization after having confirmed that right circularly polarized x-rays change the sign of the XRMR effect.
The circular polarization was realized by two 600\,$\mu$m thick diamond plates at the eight-wave plate condition mounted in series.

From the collected data the non-magnetic reflectivity $I=\frac{I_\mathrm{+}+I_\mathrm{-}}{2}$ and the asymmetry ratio $\Delta I=\frac{I_\mathrm{+}-I_\mathrm{-}}{I_\mathrm{+}+I_\mathrm{-}}$ can be determined, with $I_\mathrm{\pm}$, the XRR intensity for positive and negative magnetic field, respectively.
The evaluation of the XRR data and the XRMR asymmetry ratios were performed with the analysis tool ReMagX\cite{ReMagX}. The fitting algorithm for the non-magnetic reflectivity data $I$ was based on the recursive Parratt algorithm\cite{Parratt54}. The roughness was modelled within a N\'{e}vot-Croce approximation\cite{NevotCroce}, which allows an analytical description of the roughness within the assumption that the derivative of the optical profile across the interface is Gaussian shaped. 
For the asymmetry ratio the fitting routine was based on the Zak matrix formalism\cite{Zak}. The tool allows to model and vary magnetooptic profiles, i.e. the spatial distribution of $\Delta \delta$ and $\Delta \beta$, while keeping the absolute values for $\delta$ and $\beta$ constant. In the matrix formalism the roughness is also treated as an optical profile with a Gaussian distribution centered at the interface\cite{Elzo12}.

{\it Ab initio} calculations were performed as reported previously in Ref.[\cite{Kuschel14}].
We calculated the L$_3$ absorption edge of a spin polarized Pt thin film and aligned it with an experimental x-ray absorption spectrum (XAS) across the edge for a Pt/Fe bilayer. 
The simulated spectrum was shifted $1.7\,$eV to higher energies to fit the experimental spectrum. 
The maximum of the absorption peak is located at a photon energy of about $11567.5\,$eV. 
From the simulation of the Pt absorption edge the dependence of the magnetooptic parameters $\Delta \delta$ and $\Delta \beta$ on the photon energy can be derived. 
The {\it ab initio} calculations show the typical behaviour of the complex refraction index close to the resonance, i.e. both $\Delta \delta$ and $\Delta \beta$ only show significant values in a small range of energies around the absorption edge. While the change in absorption $\Delta \beta$ is positive with a maximum at about the peak of absorption, the variation of the dispersion $\Delta \delta$ crosses zero around the absorption edge. 
However, the maximum in $\Delta \beta$ is slightly shifted to lower energies with respect to the whiteline of the absorption spectrum.
For Pt it is well known, that the maximum of the magnetic dichroism is located slightly below the absorption edge\cite{Schuetz90, Geissler01, Kuschel14}.
Therefore, we expect the XRMR asymmetry ratio to be most pronounced for photon energies in this range.

\section{Results and discussion}
\begin{figure}[t]
	\centering
		\includegraphics[width=8.5cm]{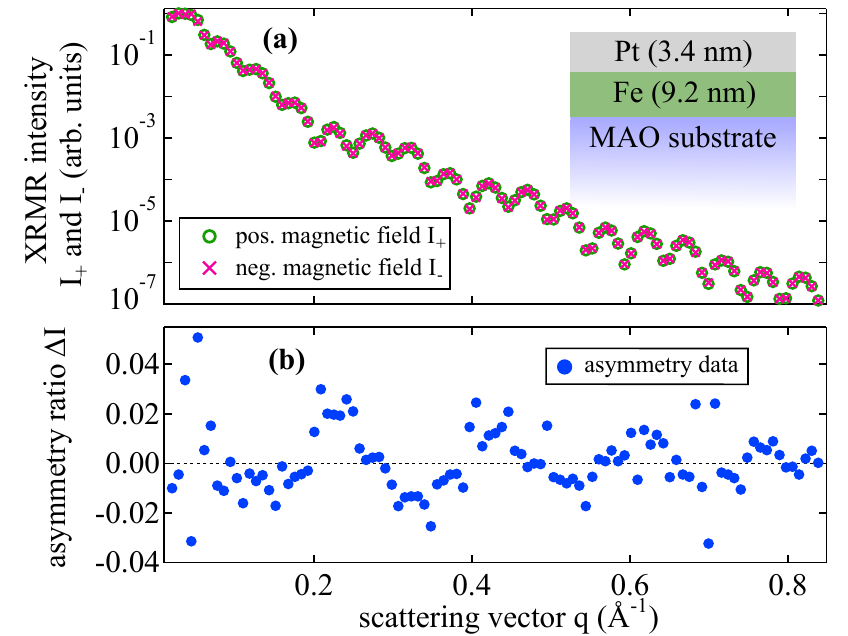}
	\caption{(Color online) (a) XRR scans for positive and negative magnetic field. (b) XRMR asymmetry ratio of the investigated Pt/Fe bilayer as derived from the XRR scans. The data were taken at a photon energy of 11567\,eV.}	
		\label{fig:XRR_Fe_Pt}
\end{figure}

Figure \ref{fig:XRR_Fe_Pt}(a) shows XRR scans of the Pt/Fe sample for positive and negative magnetic field at a photon energy of $11567\,$eV. The curves are denoted as $I_{\pm}$ for positive and negative magnetic field, respectively. 
The XRR intensities are plotted against the scattering vector $q=2\,k\,\mathrm{sin}(\theta)$, where $k$ is the wavenumber and $\theta$ is the angle of incidence with respect to the interface plane. 
The XRR curve mainly shows oscillations due to the thicker Fe film. Weaker oscillations of larger periodicity caused by the thinner Pt film on top of the supporting Fe film are superposed to these strong oscillations but are visible only weakly. From the beating effect one can estimate that the ratio of Fe to Pt film thickness is rougly 1:3.
By fitting the average of the XRR curves (not shown) we obtain the thickness (inset of \ref{fig:XRR_Fe_Pt}(a)) and roughness, and the optical constants $\delta$ and $\beta$ of the layers. We find a Fe thickness of $9.2\,$nm with a roughness of about $0.3\,$nm and a Pt thickness of $3.4\,$nm with a roughness of about $0.3\,$nm. 
Note that the sketched interfaces only indicate the center of the transition from one material to the other. The width of the transition is determined by the interface roughness.
Therefore, the depth profile of the magnetooptic parameters strongly depends on the roughness.
Calculating the XRMR asymmetry ratio, as introduced above, allows identifying small deviations in the XRR curves $I_+$ and $I_-$. 
The derived XRMR asymmetry ratio is displayed in \ref{fig:XRR_Fe_Pt}(b). Here, pronounced oscillations are visible with an amplitude of about $2\%$ indicating an induced spin polarization in the Pt.

\subsection{Magnetooptic profiles}\label{sec:models}
\begin{figure}[t]
	\centering
		\includegraphics[width=8.5cm]{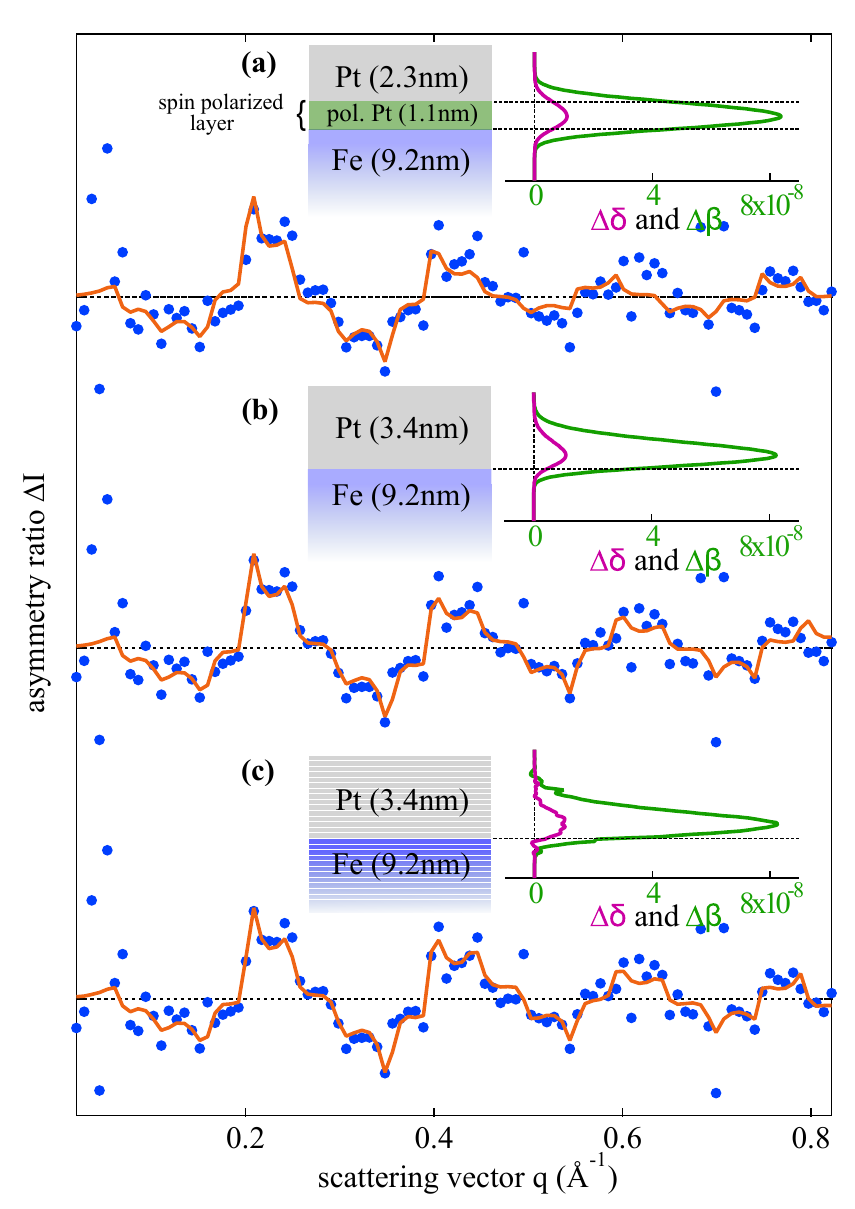}
	\caption{(Color online) The XRMR asymmetry data was fitted using three different approaches, i.e. (a) insertion of a thin spin polarized layer in the Pt, (b) convolution of the Pt/Fe interface roughness with a Gaussian shaped magnetooptic profile, and (c) an adaptive layer segmentation with separate spin polarized layers. For each model the corresponding magnetooptic profile is displayed in the inset.}	
	\label{fig:XRMR_models}
\end{figure}

In order to extract the spin polarization of the Pt from the data, the asymmetry ratio has to be fitted by implementing the change in the optical constants as magnetooptic parameters $\Delta \delta$ and $\Delta \beta$ with a certain spatial distribution across the interface.
In Fig.\ref{fig:XRMR_models} we present an overview of three different approaches we used to model the magnetooptic profiles.

In the first approach, depicted in Fig.\ref{fig:XRMR_models}(a), we divide the Pt film into an unpolarized layer and an additional fully spin polarized layer with a finite roughness. The resulting magnetooptic profiles for $\Delta \delta$ and $\Delta \beta$ arise from the convolution of the magnetooptic parameters with the interface roughnesses between the two separate Pt layers and between the spin polarized Pt and the Fe film. 
We obtained the illustrated distribution of $\Delta \delta$ and $\Delta \beta$ by fitting the thickness, the roughness, and the magnetooptic constants of the polarized Pt layer, while keeping the total Pt thickness constant. 
Though, in order to limit the set of fit parameters, the $\Delta \beta$/$\Delta \delta$ ratio was fixed to 7.6, as derived from the optical data from the {\it ab initio} calculations at this photon energy. Note that the $\Delta \beta$/$\Delta \delta$ ratio of 7.6 corresponds to the photon energy 11567\,eV, while in an earlier work we carried out our measurements at 11567.5\,eV and consequently found a $\Delta \beta$/$\Delta \delta$ ratio of 3.4\cite{Kuschel14}.
The effective spin polarized film thickness appears to be about $(1.1 \pm 0.1)\,$nm in this model, while the roughness derived by the fitting is about 0.4\,nm. 

In Fig.\ref{fig:XRMR_models}(b) the second model is illustrated. Instead of including an additional Pt layer, the magnetooptic profiles are estimated by a convolution of the interface roughness with a Gaussian shaped profile. In this approach the median, the variance and the amplitudes, $\Delta \delta$ and $\Delta \beta$ of the Gauss distribution are fit parameters. Again, $\Delta \beta$/$\Delta \delta$ was set to 7.6. This approach results in a fit curve very close to the experimental data and a realistic magnetooptic profile. The FWHM of the magnetooptic profiles, i.e. the effective width of the spin polarized Pt, is about $(1.2 \pm 0.1)\,$nm. 
This approach was also used by Br\"uck et al. to model the asymmetry ratio of spin polarized Mn in Fe/MnPd bilayers\cite{Brueck10}.

The third model is shown in Fig.\ref{fig:XRMR_models}(c). Here we use a layer segmentation to simulate the spatial distribution of $\Delta \delta$ and $\Delta \beta$.
Starting with the profile from the second model, the system is divided into thin sublayers with a distinct thickness and zero roughness. In this case the roughness is modelled by a gradual transition of the optical constants from one material to the other. Each of the sublayers can exhibit a finite value for $\Delta \delta$ and $\Delta \beta$, which are fitted to obtain the best simulation for the experimental data. This allows a high degree of freedom for the shape of the magnetooptic profiles.
Again, the resulting FWHM of the magnetooptic profiles is about $(1.2 \pm 0.1)\,$nm.

Evaluating the three approaches we found distinct differences in the simulated and experimental data.
The first model of a spin polarized Pt interlayer shows the largest deviations between simulation and experiment. 
This is reflected in a larger $\chi^2$ goodness of fit. 
Also, since the Zak matrix formalism treats the roughness as a Gaussian distribution around the interface, problems in the fit with this model may arise if the roughness is in the same range or larger than the interlayer thickness. 
The second model leads to a convincing fit but shows some slight deviations for high values of the scattering vector q, while the third model describes the data very well for the entire range. However, this model consumes much more computing-time and works with a large set of correlated fitting parameters.
Within the experiment the second model simulates the data with the smallest number of parameters and results in a plausible fit; therefore, this model was chosen for the data evaluation, analogous to our previous work\cite{Kuschel14}.

\subsection{Photon energy dependence}
\begin{figure}[t]
	\centering
		\includegraphics[width=8.5cm]{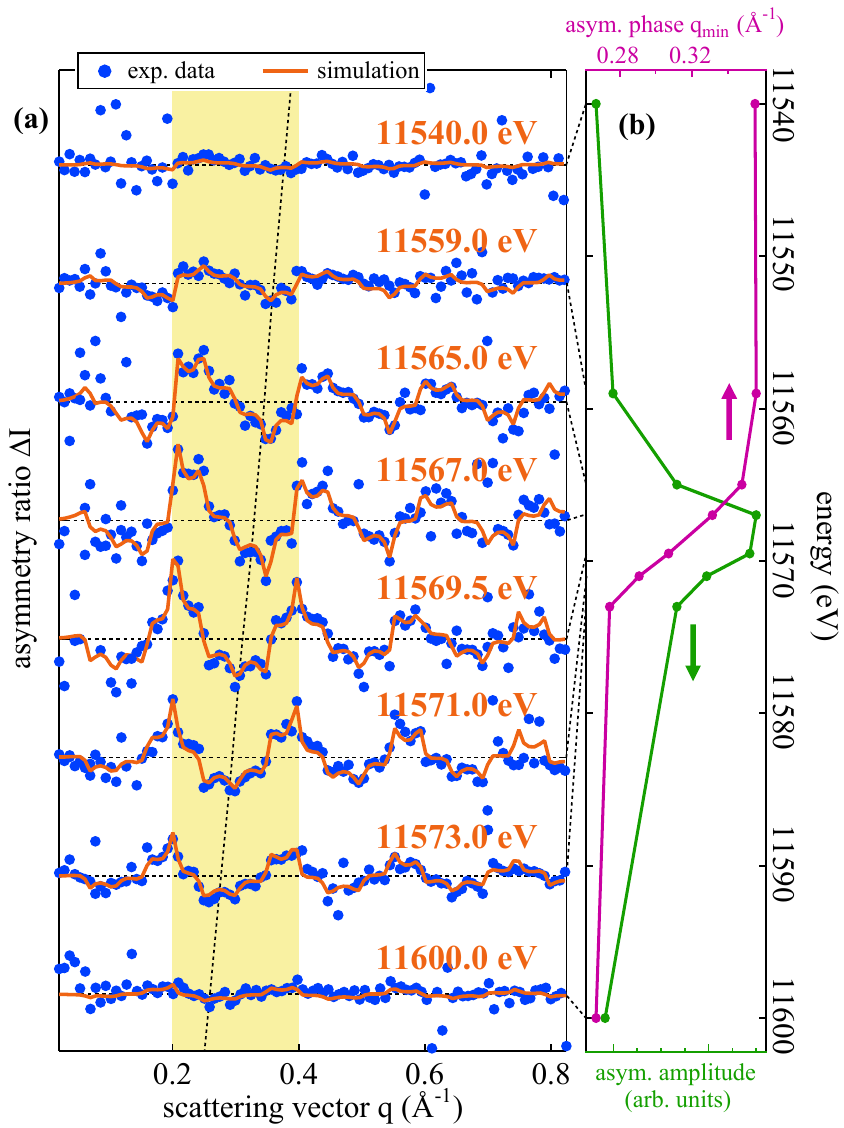}
	\caption{(Color online) (a) XRMR asymmetries and fits for Pt(3.4\,nm)/Fe(9.2\,nm) at different photon energies around the Pt L$_3$ absorption edge. The region between q$=0.2\,\text{\AA}^{-1}$ and q$=0.4\,\text{\AA}^{-1}$ is highlighted to illustrate the phase shift. The positions of the minima are further traced by a dashed line.
	(b) Asymmetry amplitude and phase shift of the oscillations, as evaluated from the raw asymmetry data.}	
	\label{fig:XRMR_fits_energy}
\end{figure}

\begin{figure}[t]
	\centering
		\includegraphics[width=8.5cm]{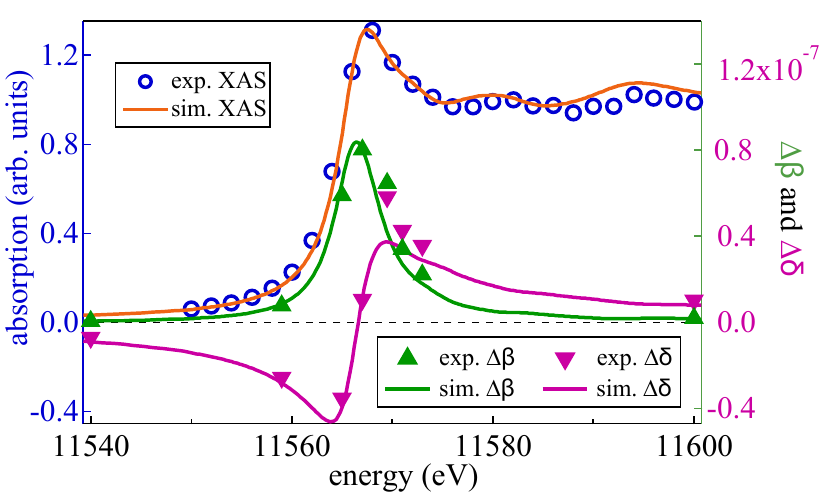}
	\caption{(Color online) Experimental and calculated energy dependent XAS spectra and corresponding magnetooptic parameters $\Delta \delta$ and $\Delta \beta$. The simulations were shifted by 1.7\,eV to higher energies to fit the experimental data correctly.}
	\label{fig:XAS}
\end{figure}

XRMR data were collected for a Pt/Fe sample in a range of photon energies around the Pt L$_3$ absorption edge between 11540\,eV and 11600\,eV (Fig.\ref{fig:XRMR_fits_energy}(a)).
Outside of this range the amplitude of the asymmetry ratio vanishes almost completely and is most pronounced for energies between 11565\,eV and 11571\,eV, close to the absorption edge. This allows to exclude any influences from other absorption edges.
The region between q$=0.2\,\text{\AA}^{-1}$ and q$=0.4\,\text{\AA}^{-1}$ is highlighted, and the dashed line marks the position of a minimum of the asymmetry ratio to illustrate the phase shift in the asymmetry ratio with respect to the photon energy. The evaluation shows, that the phase shift is mainly determined by the change of $\Delta \delta$, while the change in amplitude is primarily governed by the variation of $\Delta \beta$ with energy. This correlation can be seen by comparing the depicted amplitude and phase shift of the asymmetry ratios in Fig.\ref{fig:XRMR_fits_energy}(b) with the simulated magnetooptic parameters $\Delta \delta$ and $\Delta \beta$, shown in Fig.\ref{fig:XAS}.

The data were fitted using the second simulation model based on a Gaussian magnetooptic profile convoluted with the interface roughness, as described in section \ref{sec:models}.
Again, prior to the asymmetry ratio, each XRR curve obtained from averaging the XRR curves measured for positive and negative field was fitted.
The structural parameters were deduced from the reflectivity data collected off resonant, at 11540\,eV, and kept constant for the remaining energies, while only the optical constants $\delta$ and $\beta$ were adjusted to the XRR scans. 
Based on these values, the spatial distribution and the quantitative values of the magnetooptic parameters were extracted from simulations of the asymmetry ratios.
As before, the ratio $\Delta \beta$/$\Delta \delta$ was fixed to the corresponding values for each energy which were acquired from the {\it ab initio} calculations for spin polarized Pt. 

To compare the experimental data and simulation results with theory, Fig.\ref{fig:XAS} also displays the results for XAS spectra from {\it ab initio} calculations and the corresponding results from fitting the experimental data.
The XAS data show a whiteline intensity of 1.31, which indicates a mostly metallic state for the Pt layer\cite{Kolobov05}.
The experimental data for the magnetooptic parameters $\Delta \delta$ and $\Delta \beta$ show the predicted behaviour.

By scaling the magnetooptic data from the {\it ab initio} calculations to the experimentally derived values for $\Delta \delta$ and $\Delta \beta$ (shown in Fig.\ref{fig:XAS}) we determined the magnetic moment induced in the Pt by magnetic proximity. We find a magnetic moment of $m_\text{Pt}=(0.43 \pm 0.10)\mu_\text{B}$ per Pt atom in a $(1.2 \pm 0.1)\,$nm thick effective layer at the interface to the Fe. This value is smaller than the value reported in an earlier publication\cite{Kuschel14} for the identical sample and for additional samples of different thicknesses. 
This discrepancy can be attributed to the fact that we derived different values for the absorption coefficient $\beta$ from the fits of the XRR curves. 
In Ref[\cite{Kuschel14}] we found significantly larger values for $\beta$ than in the present experiments, which results in a larger absolute value of the magnetooptic parameter $\Delta \beta$ for the same amplitude of the asymmetry ratios, and therefore in a larger moment. 
We can only speculate on the origin of the different absorption coefficients. 
Since the experiments were performed six months apart, one possible reason are aging effects like oxidization and interdiffusion altering the surface and the interfaces of the investigated sample.

However, these quantitative values for the induced moment and the width of the spin polarized volume in Pt are consistent with previous reports on Pt/Fe investigated by XMCD\cite{Antel99} (see Fig.\ref{fig:Proximity_Pt}).

\subsection{FM material dependence}
\begin{figure}[t]
	\centering
		\includegraphics[width=8.5cm]{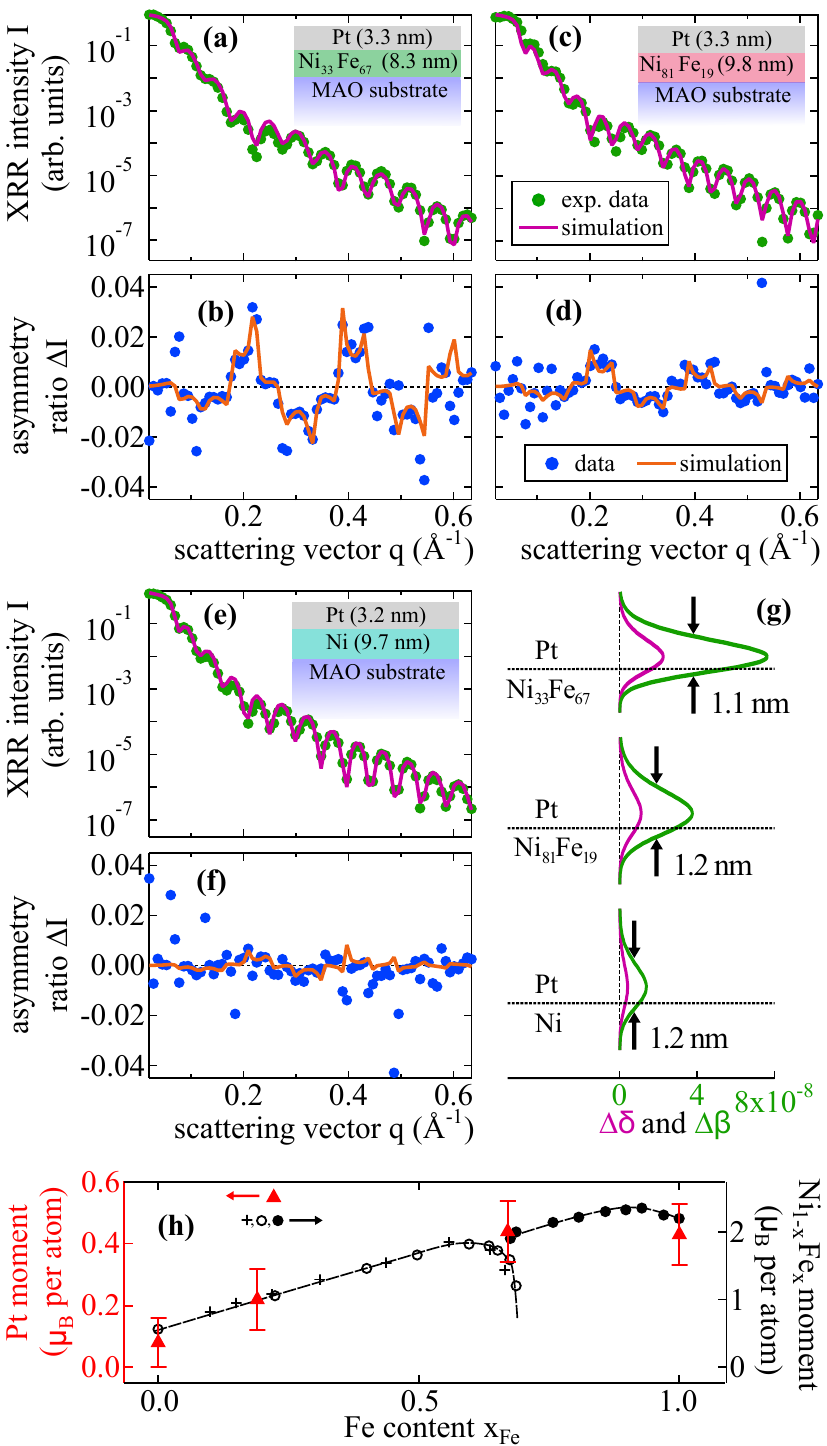}
	\caption{(Color online) XRR scans and XRMR asymmetry ratios for different Pt/FM bilayer combinations. (a),(b) Pt/Ni$_{33}$Fe$_{67}$, (c),(d) Pt/Ni\(_{81}\)Fe\(_{19}\), (e),(f) Pt/Ni. The corresponding magnetooptic profiles are displayed in (g). (h) shows the dependency of the Pt spin polarization on the Fe content in the underlying FM (red triangles). The black circles and crosses are reproduced from Ref.[\cite{{Crangle63, Kondorskii52}}] and display the magnetic bulk moment of Ni$_{1-x}$Fe$_x$. The solid circles\cite{Crangle63} represent data for a crystallization in bcc structure and the open circles\cite{Crangle63} and crosses\cite{Kondorskii52} represent data from samples with fcc structure.}
	\label{fig:XRMR_fits_materials}
\end{figure}

In Fig.\ref{fig:XRMR_fits_materials} we present XRMR investigations on different Pt/FM bilayer systems, collected at a photon energy of 11567.5\,eV.
In addition to the measurements on Pt/Fe we implemented Ni$_{33}$Fe$_{67}$, Ni\(_{81}\)Fe\(_{19}\), and Ni as the ferromagnetic material. For each layer combination the XRR scans and the asymmetry ratios are shown in Fig.\ref{fig:XRMR_fits_materials}(a)-(f). 
Similar to the Pt/Fe sample, the film thicknesses of the FMs are in the range of $8.3\,$nm to $9.8\,$nm, while the Pt had a thickness between $3.2\,$nm and $3.3\,$nm (see insets).
Both, the Pt/Ni$_{33}$Fe$_{67}$ and the Pt/Ni\(_{81}\)Fe\(_{19}\) sample show a finite amplitude in the asymmetry ratios (Fig. \ref{fig:XRMR_fits_materials}(b) and (d)), implying a significant amount of spin polarized Pt. The induced magnetic moments of the Pt as determined from the experimental data for Pt on Ni$_{33}$Fe$_{67}$ and on Ni\(_{81}\)Fe\(_{19}\) are $m_\text{Pt/Ni\(_{33}\)Fe\(_{67}\)}=(0.44 \pm 0.10)\mu_\text{B}$ and $m_\text{Pt/Ni\(_{81}\)Fe\(_{19}\)}=(0.22 \pm 0.10)\mu_\text{B}$ per atom, respectively.

The XRMR data for the Pt/Ni bilayer show only weak oscillations in the asymmetry ratio (Fig. \ref{fig:XRMR_fits_materials}(d)) corresponding to a moment of about $(0.08\pm0.08)\mu_\text{B}$ per atom. 
The presented asymmetry ratio was recorded in only one measurement, which explains the comparably small signal-to-noise ratio. 
However, we were able to show that this can be improved significantly by averaging over a greater number of measurements\cite{Kuschel14}. 
Due to the small signal-to-noise ratio the simulation was done with fixed parameters as given in Fig.\ref{fig:XRMR_fits_materials}(g), in order to obtain an upper limit for the Pt polarization. 
In a previous work Wilhelm et al. have observed an induced moment of up to $0.29\mu_\text{B}$ in Pt/Ni multilayers with Pt thicknesses of only 2 monolayers \cite{Wilhelm2000} (see Fig.\ref{fig:Proximity_Pt}). This value is considerably larger than the moment found in our Pt/Ni bilayer. However, the data from Wilhelm et al. were recorded at 10K and are therefore likely to result in a higher moment compared to RT measurements.

The magnetooptic profiles for the three sample systems are shown in Fig.\ref{fig:XRMR_fits_materials}(g). Again, the effective width of the spin polarized volume is in a range between 1.1\,nm and 1.2\,nm for the three samples. 
Fig.\ref{fig:XRMR_fits_materials}(h) shows the induced Pt moment in dependence on the Fe content (red triangles).
The induced spin polarization scales with the amount of Fe in the FM. Data for the magnetic bulk moments of various Ni$_{1-x}$Fe$_x$ compounds (reproduced from Ref.[\cite{Crangle63, Kondorskii52}]) show a comparable dependence on the Fe content x$_\text{Fe}$. Both the induced Pt moments and the bulk moments of the Ni$_{1-x}$Fe$_x$ compounds decrease significantly for a vanishing Fe content in the layer. 
The presented results suggest, that the strength of the magnetic coupling between the two layers depends on the magnitude of the magnetic moment in the FM. A similar result was found by Wilhelm et al. for Pt/Ni and Pt/Co\cite{Wilhelm03} samples.

In general, it is well understood that magnetic proximity effects are mainly governed both by band hybridization at the interface between the NM and the FM and exchange interactions across the interface\cite{Cox79, Bluegel89}. 
However, Altbir et al. \cite{Altbir89} stated that for a weak FM, i.e. a small splitting between the spin-up and spin-down bands, the expansion of the magnetization into the NM is very small even for the atomic layers closest to the interface.
This is consistent with our findings of a reduced Pt moment in Pt/Ni bilayers. 
Nevertheless, a more detailed theoretical description of the underlying coupling mechanism in the investigated systems remains pending.

\section{Conclusion}
We performed XRMR characterization of Pt/FM bilayer systems to investigate the Pt interface spin polarization, when adjacent to different ferromagnetic materials. 
The XRMR asymmetries were quantitatively analysed by fitting the magnetooptic parameters and comparing the experimental findings with {\it ab initio} calculations. 
Different magnetooptic profiles were applied to the asymmetry ratio of a Pt/Fe bilayer and evaluated in order to find the best fitting model. A convolution of the interface roughness with a Gaussian shaped profile yielded the best results. 

The XRMR asymmetry ratios were taken at a varying photon energy for Pt/Fe and the results were compared to {\it ab initio} simulations for spin polarized Pt.
The dependence of the extracted magnetooptic parameters $\Delta \delta$ and $\Delta \beta$ on the photon energy shows good agreement with the calculated behaviour close to the absorption edge and allows determining the induced Pt moment. 
For the magnetic moment in the Pt/Fe bilayers we found $m_\text{Pt}=(0.43 \pm 0.10)\mu_\text{B}$ per Pt atom in the spin polarized volume at the interface. 
The reduced moment obtained here compared to an earlier work\cite{Kuschel14} is attributed to aging effects in the investigated sample.

Additionally, we studied samples with different FM layers.
We find a correlation of the Pt spin polarization and the Fe content of the adjacent ferromagnet in Pt/Fe, Pt/Ni$_{33}$Fe$_{67}$, Pt/Ni\(_{81}\)Fe\(_{19}\), and Pt/Ni bilayers. We observe a reduction of the magnetic moment of Pt in proximity to Ni compared to the Pt/Fe bilayer. 

\section*{Acknowledgments}
The authors gratefully acknowledge financial support by the Deutsche Forschungsgemeinschaft (DFG) within the Priority Program Spin Caloric Transport (SPP 1538).
Further they are grateful for the opportunity to work at beamline P09, PETRA III at the Deutsche Elektronen Synchrotron (DESY) and thank for technical support by David Reuther and Sonia Francoual.
They also thank Sebastian Macke for providing software support of the fitting tool ReMagX.  
The authors further thank Gerhard G\"otz and Daniel Meier for performing characterization measurements of the investigated samples.

\end{document}